# Optimizing and extending light-sculpting microscopy for fast functional imaging in neuroscience

Peter Rupprecht,[1-4] Robert Prevedel,[1-3] Florian Groessl,[1] Wulf E. Haubensak[1] and Alipasha Vaziri[1-3,*]

[1]*Research Institute of Molecular Pathology, Vienna, Austria.*
[2]*Max F. Perutz Laboratories, University of Vienna, Vienna, Austria.*
[3]*Research Platform Quantum Phenomena & Nanoscale Biological Systems (QuNaBioS), University of Vienna, Vienna, Austria.*
[4]*Current address: Friedrich Miescher Institute, Basel, Switzerland*
[*]*alipasha.vaziri@imp.ac.at*

**Abstract:** A number of questions in systems biology such as understanding how dynamics of neuronal networks are related to brain function require the ability to capture the functional dynamics of large cellular populations at high speed. Recently, this has driven the development of a number of parallel and high speed imaging techniques such as light-sculpting microscopy, which has been used to capture neuronal dynamics at the whole brain and single cell level in small model organism. However, the broader applicability of light-sculpting microcopy is limited by the size of volumes for which high speed imaging can be obtained and scattering in brain tissue. Here, we present strategies for optimizing the present tradeoffs in light-sculpting microscopy. Various scanning modalities in light-sculpting microscopy are theoretically and experimentally evaluated, and strategies to maximize the obtainable volume speeds, and depth penetration in brain tissue using different laser systems are provided. Design-choices, important parameters and their trade-offs are experimentally demonstrated by performing calcium-imaging in acute mouse-brain slices. We further show that synchronization of line-scanning techniques with rolling-shutter read-out of the camera can reduce scattering effects and enhance image contrast at depth.

OCIS codes: (110.0110) Imaging systems; (110.0180) Microscopy; (180.2520) Fluorescence microscopy; (170.1420) Biology; (170.3660); Light propagation in tissues; (170.3880) Medical and biological imaging; (170.5810) Scanning microscopy; (180.6900) Three-dimensional microscopy; (170.2655) Functional monitoring and imaging; Imaging through turbid media; (320.0320) Ultrafast optics; (320.5540) Pulse shaping.

## 1. Introduction

Currently there are major efforts in different areas of biology and neuroscience to image volumes up to the level of whole organisms and brains at high spatial and temporal resolution [1-8]. In neuroscience in particular the goal of understanding how the dynamics of the neuronal network is linked to brain function and behavior requires recording from neurons spread over large populations on physiologically relevant time-scales. Over the last decade, the combination of two-photon microscopy and genetically encoded calcium indicators (GECI) has emerged as an indispensable tool for optical readout of neuronal activity. GECIs such as GCaMP [9] are widely used for efficient and cell type specific labeling of neurons and allow mapping the intracellular calcium levels, a proxy for neuronal activity, into changes of a fluorescence signal. On the hardware side, two-photon point scanning microscopy has evolved as the gold standard for imaging as it provides the necessary lateral as well as axial resolution, signal to background ratio and improved depth penetration in biological tissue [10]. However, the above advantages of standard two-photon scanning microscopy come at the cost that a diffraction limited excitation spot has to be scanned in the lateral plane, leading to low temporal resolution. Thus, it has remained a challenge to functionally image large volumes at single cell resolution and physiological time scales. While some of the currently available techniques address some aspects of the above requirements, it has remained challenging for a single technique to fulfill all of them at the same time. Given that in almost all current imaging techniques, there is a more or less straightforward tradeoff between the imaging speed, spatial resolution, volume size and shape, signal strength and imaging duration, it is important to choose these parameters appropriately for the biological question in mind. Here we discuss the details of such tradeoffs for the emerging technique of light sculpting based on temporal focusing and provide guidelines for optimally choosing various imaging parameters under different conditions and available laser sources. In addition we provide a simple strategy for minimizing the effects of tissue light scattering in light-sculpting microscopy and quantify its performance in $Ca^{2+}$-imaging experiments.

## 2. State-of-the-art in functional imaging

The current diffraction limited two-photon scanning approaches have varying performance. For concreteness, assuming a typical 512x512 pixel plane, standard galvanometric point-scanning with a scanning frequency of 1 kHz yields a frame rate of ~4 Hz for bidirectional scanning. Strategies to overcome this speed limit include random access scanning using acousto-optical deflectors (AODs) [4-6] a method designed for imaging with rates of up to ~50kHz/N, where N is the number of points. As a second possibility, fast plane scanning using AODs [11] or resonant scanners [12-14] can increase the frame rate considerably. For example, a typical 8 kHz resonant scanner would achieve video rate (30 Hz) in bidirectional scanning mode for a 512x512 pixel image. Such an increase in scan speed always has to be accompanied by an increase in illumination intensity to maintain a useful signal-to-noise level, but is limited by the number of fluorophores in a sample for laser scanning microscopy [15].

Alternatively, light-sculpting techniques based on temporal focusing (TeFo) [16-18] have been put forward that circumvent the scanning speed limitations associated with such point-like scanning techniques [3, 19-23]. In addition, the parallel acquisition scheme using a camera leads to dramatically increased pixel dwell time compared to fast point scanning with the same frame rate. This reduces photodamage, bleaching and limitations due to saturation of the fluorophores and thereby increases the effective photon yield compared to laser scanning microscopy [1, 15]. In temporal focusing, the spectrum of a femtosecond pulsed laser is spatially dispersed by a grating and imaged onto the sample by a telescope, consisting of a lens and the objective lens. Thereby the frequency components of the laser pulse are spatially dispersed everywhere but at the focus of the objective lens. This temporally stretches the pulse, leads to a lower peak pulse intensity and thus lowers the two-photon excitation probability outside the focal region. Since in this scheme the axial localization of excitation is mainly achieved by controlling the dispersion of the pulse in the sample, a laterally wide, but axially confined excitation pattern can be produced (so-called wide-field temporal focusing – WF-TeFo – configuration [3, 20]).

However, such a wide-field approach to volumetric imaging also poses two main limitations. First, applying wide-field temporal focusing to $Ca^{2+}$-imaging means that a large area, covering at least tens to hundreds of neurons needs to be excited via two-photon excitation within the exposure time allocated to each image plane. This typically requires costly regenerative amplifier systems, which can provide the necessary high peak powers [3]. However, this limits the broader use of this approach in biology and neuroscience due to the cost- and maintenance intensive amplified lasers and the rather small obtainable area of excitation. To overcome this problem, a smaller excitation area element can be sculpted and scanned over the imaging plane to cover the desired area [18, 23]. Because of the nonlinear nature of two photon excitation, the overall fluorescence yield for a given imaging area, frame exposure time and laser power can be enhanced when only a small fraction of the area is excited at a given time, and the excitation area is scanned over the whole imaging area. For this work, we use two such alternative configurations: one where a temporally focused line is scanned in one dimension (1D-LS) and another where small (~4-8 μm diameter) temporally focused spots are scanned in two dimensions using spiral scanning (2D-SS). As mentioned above, we will provide guidance for finding the most appropriate choice between these different modalities for given sample requirements in the context of $Ca^{2+}$-imaging. The design and parameter choices are based on some of the currently available laser sources and their respective optical performance, the desired field-of-view (FOV) or lateral size of excitation and the required temporal resolution. Our findings are corroborated experimentally by calcium imaging in acute mouse brain slices, a simple model for benchmarking novel physiological technology.

Second, while two-photon excitation reduces the effects of scattering for the incoming light, scatter of the emitted fluorescence results in cross-talk of neighboring pixels in the parallel readout via 2D sensors such as sCMOS or EMCCD cameras and therefore limits the obtainable imaging depth. For non-temporal focusing line-scanning configurations, methods have been developed that de-scan the fluorescence of the illuminated line onto a static 1D confocal pinhole that rejects out-of-focus and background light [24-26]. A simpler technique has recently been proposed [27] that exploits the synchronization of line-shaped illumination of a sample with readout of modern sCMOS cameras. sCMOS cameras normally do not expose the whole sensor at once ('global shutter'), but only a given number of pixel rows (slit) at a time that move over the sensor ('rolling shutter'). If the size of the slit is reduced and synchronized to the illuminating line, this "virtual" confocal slit reduces the effects of scattering in the direction orthogonal to the line. The working principle of this rolling-shutter detection in combination with one-photon light-sheet illumination was experimentally shown with fluorescent beads and fixed biological samples [28]. Here, we explored the potential of this technique for line-scan temporal focusing and in the context of high-speed calcium imaging. We first used Monte Carlo simulations to identify the magnitude of the expected scattering effects in typical brain tissue, and investigated to what extent scattered light can be rejected using the rolling shutter detection. Consistent with our theoretical findings, and Ref. [28], we then experimentally demonstrate the expected enhancement in image contrast and spatial resolution when calcium imaging acute mouse brain slices.

## 3. Experimental realization

Our setup is schematically depicted in Fig. 1A-B. As in any microscope setup, the choice of imaging optics determines the field of view in TeFo microscopy. The microscope optical design has been optimized for the 20x 1.0NA objective employed in our experiments. The setup includes an xy-galvometric mirror set (6215H, Cambridge Technology) that allows for scanning in 2D. The scanning mode can be easily switched by replacing the scan lens directly after the scan mirrors in the beam path (Fig. 1B). Employing no scan lens results in the wide-field configuration, while a cylindrical scan lens ($f_{cyl}$) shapes a line in the sample. The spiral-scan configuration is achieved by replacing the cylindrical lens by a spherical lens. We use two different excitation laser sources in our study, a standard Titanium Sapphire (Ti:Sa) laser (Coherent Inc, Chameleon Ultra II) and a single pass regenerative amplifier (Coherent Inc, Legend Elite) in combination with an optical parametric oscillator (Light Conversion, TOPAS-C). The objective lens (Olympus XLUMPLFLN 20x 1.0NA) together with the 2" lens ($f = 1000$ mm) formed the telescope which imaged the illuminated spot from the grating onto the specimen plane. In order to achieve a 200 µm-wide FOV in the sample with the demagnification of ~111 provided by the telescope, a ~22 mm wide spot had to illuminate the grating. The combination of the above lens and the grating period (800 lines/mm) was chosen such that the FWHM of the spatially spread pulse was covering the back focal aperture (BFA) of the objective. We note here that the size of our grating (30x30mm), along with the 2" optics employed, limited the largest possible FOV to <250 µm in our setup, due to aperture clipping effects. Larger diameter optics and/or smaller demagnification ratios of the telescope combined with gratings of higher groove density to maintain sufficient spatial dispersion could in principle further increase the FOV.

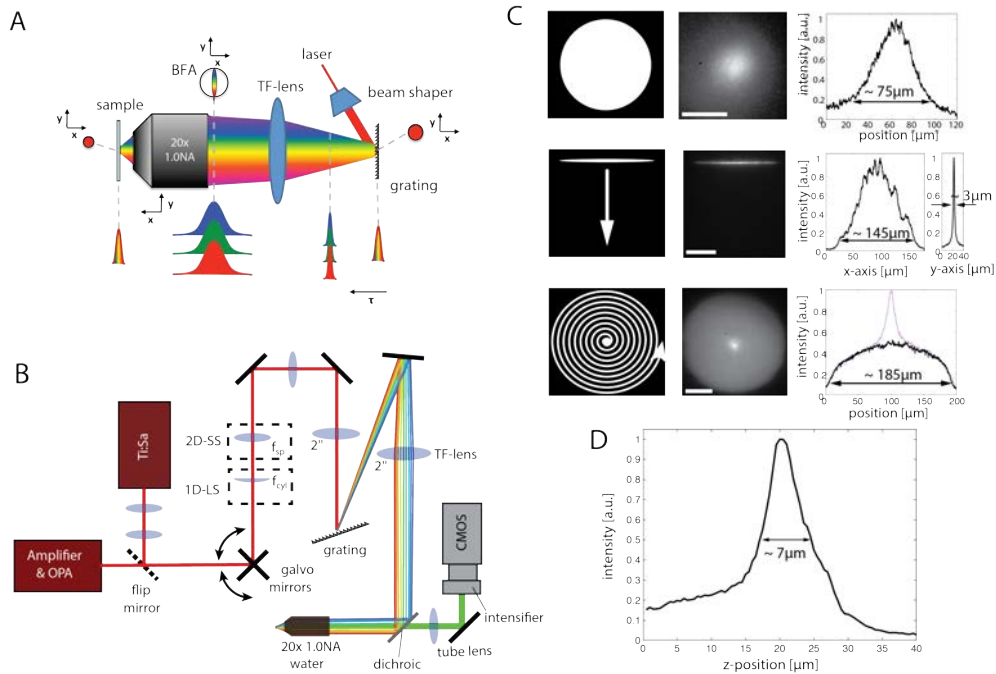

Fig. 1. Experimental setup and various modalities of light sculpting microscopy. **A.** Principle of temporal focusing. The grating disperses the laser pulse in its spectral components, which are refocused by a telescope in time and space formed by the temporal focusing lens (TF-lens) and the objective at the image plane in the sample. The dashed lines illustrate the spatial and temporal dispersion of the pulse at various points along the axial direction **B.** Schematic of the setup including its various modalities. The laser beam, from either a Ti:Sa source or an amplified system, is expanded via lenses and directed towards a pair of galvo mirrors, after which, depending on the scanning modality either no lens (wide-field), a cylindrical scan lens ($f_{cyl}$ – line-scan 1D-LS) or a spherical scan lens ($f_{sp}$ – spiral-scan 2D-SS) are employed. **C.** Left panel illustrates the schematics of various excitation modalities, including wide-field (top), line-scan 1D-LS (middle) and spiral-scan 2D-SS (bottom). Note that in the spiral scan configuration the small spot (diameter d~4-8 µm) being scanned is also a temporally focused disc. The middle panel shows the resulting excitation pattern taken with a fluorescent plastic slide, with cross sections along the x- and y-axis shown in the right panel. **D.** A typical axial scan of a single, 500 nm fluorescent bead in the spiral scan configuration, indicating the axial resolution. The asymmetry is due to background fluorescence. Scale bars are 50 μm in C.

## 4. Comparison of scanning modalities in TeFo microscopy

The optimal scanning scheme depends on the requirements on volume size and shape, temporal and spatial resolution as well as (available) signal strength. In practice, additionally the parameters of the commercially available laser sources as well as the ease of operation and costs put a significant constraint on the choice of the excitation strategy. Ultra-fast Titanium Sapphire (Ti:Sa) oscillators with repetition rates ~80 MHz are widely used in neuroscience labs, due to their high average power (> 1W), broad wavelength tunability (~700-1000 nm) and more recently due to their hands-off user-friendly operation. However, the need for fast [3], and deep [29] functional 3D imaging has often necessitated the use of laser sources with higher pulse peak powers, such as provided by regenerative amplifier. In this work, we make use of both of these systems in combination with the different temporal focusing excitation modalities described below and evaluate their respective strengths and the involved trade-offs. The two laser systems used are a Ti:Sa oscillator (Coherent Chameleon, repetition rate 80MHz, pulse length 130fs) and a regenerative amplifier system (Coherent Legend Elite, 10kHz, 130fs) in combination with an OPA for wavelength tuning (Light Conversion, TOPAS-C). These systems delivered a maximum average power of 250 mW (3.1 nJ per pulse) and 25 mW (2.5 μJ per pulse), respectively, to the sample plane at 920 nm, the peak two-photon absorption wavelength of the calcium indicator GCaMP [9]. These values were also used for our theoretical calculations below.

To compare different TeFo excitation modalities, one has to evaluate the expected fluorescent signal. The number of absorbed photons per fluorophore, $N_a$, and therefore the fluorescence signal in two-photon excitation via a pulsed laser source is proportional to [30]:

$$N_a \sim \frac{P_0^2}{f\,\tau}\left(\frac{\lambda}{A}\right)^2 \Delta t \quad , \quad (1)$$

with $P_0$ the average laser power at the sample plane, $f$ the laser's pulse repetition rate, $\tau$ the pulse length, $\lambda$ the central wavelength, $A$ the excited area (*e.g.* the area of the line or of the small disc for LS line- and SS spiral-scanning, respectively) at the sample and $\Delta t$ the dwell (or exposure) time. A key aspect of this equation is the quadratic dependence of the number of absorbed photons on $A$ and the linear proportionality in $\Delta t$. For example, if the area element $A$ is reduced by a factor of 10, the dwell time $\Delta t$ on a given location needs to be reduced by the same factor to maintain the same imaging frame rate. However, since $N_a \sim \Delta t / A^2$, the fluorescence yield will be increased by a factor 10 in this example, thereby allowing further decrease in dwell time and hence an increase in the frame rate.

To illustrate the trade-offs of $A$ and $\Delta t$ for different configurations, we assume in the following a FOV of a given size, i.e. $A_{FOV} = 200x200$ μm and an effective plane exposure time of $t_{exp} = 10\text{ ms}$, which is necessary to image the whole FOV. For the wide-field configuration (WF-TeFo), no lateral scanning is required (Fig. 1C-top). The area $A$ equals the whole FOV ($A = A_{FOV} = 200x200$ μm$^2$) and $\Delta t$ equals the exposure time $t_{exp}$ of a plane (10 ms). For the line-scan configuration, the line has to be scanned in one dimension in order to cover the whole FOV (Fig. 1C-middle). $A$ is given by the area of the line (*e.g.* ~$200 \times 3$ μm² for a line width of 3 μm) and $\Delta t$ by the dwell time on each line ($\Delta t = t_{exp} \cdot A / A_{FOV}$). For the spiral-scan configuration, a small TeFo spot ($d = 4-8$ μm diameter for optimal performance) is scanned with constant velocity on the trajectory of a circular involute in order to homogeneously fill the FOV (Fig. 1C-bottom). In this case $A = \pi d^2 / 4$ with $d = 4-8$ μm diameter of the TeFo spot, and again the dwell time $\Delta t$ is given by $\Delta t = t_{exp} \cdot A / A_{FOV}$. The increase in fluorescence signal $N_a$ compared to the widefield configuration is therefore, in all cases, given by the factor $A_{FOV} / A$.

For each configuration, the parameters have to be chosen such that $N_a$ attains at least a signal level above background and noise levels of the detection system. A certain signal to noise ratio (SNR) is further necessary for this signal to be useful for processing and meaningful interpretation. As reference, we chose an empirical value of $N_a \sim 3 \cdot 10^8$ that typically gives sufficient SNR (typically >2) for Ca$^{2+}$-imaging with GCaMP5/6 with our setup, consistent with our previous study [3]. We have empirically found this value for $N_a$ to be a good reference point and used it to compare the expected and the experimental performance for the two different laser light sources and the three scanning modalities described above.

Both our theoretical estimates and experimental results are shown in Fig. 2. Fig. 2A shows the dependence of the expected fluorescence signal ($N_a$) on the FOV for different temporal focusing excitation modalities and laser systems. The two laser systems in our experiment can, in principle, be used in conjunction with all three scanning modalities. However, in most functional experiments, the comparably low peak power of the Ti:Sa is prohibitive for wide-field illumination (*i.e.* > 20 μm, Fig. 2A). On the other hand, the lower repetition rate of the amplifier system makes spiral-scanning unpractical due to the resulting inhomogeneous illumination, although an amplifier system at higher repetition rates (e.g. as used in Ref. [23]) may prove advantageous for this application. Additionally, the maximum scan speed of the galvanometric mirrors limits the number of spiral loops within a given plane exposure time and therefore the achievable maximum FOV for a given temporally focused disc size (dashed lines in Fig. 2A). Fig. 2B shows isolines of the expected signal when trading-off $A_{FOV}$ and exposure time $t_{exp}$. For example, if a plane exposure time of $t_{exp} = 10$ ms is required, the plot provides a lookup table for a typical FOV size that can be achieved with each configuration. For the sCMOS camera (Andor Neo) that we used, a plane exposure time of 10 ms translates into an effective frame rate of 75-80 Hz for a ~170x170 μm (512x512 pixel) image.

Overall, we can draw the following conclusions: At a constant frame exposure time ($t_{exp} = 10$ ms) and with an available average power of 25 mW at the sample plane for a 10 kHz amplified system, the wide-field temporal focusing (WF-TeFo) approach provides sufficient signal for small (<75 μm diameter) FOVs, as previously demonstrated [3], but this configuration scales rather unfavorably for larger FOVs. An interesting and viable alternative to amplified systems when only a Ti:Sa oscillator is available, is provided by the spiral-scanning (2D-SS TeFo) approach of a small temporally focused spot. For the same level of collected fluorescence signal it allows excitation of larger (up to ~150 μm) FOVs compared to WF-TeFo. The line-scanning approach (1D-LS TeFo), when combined with an amplified laser system, provides the highest signal levels for the same exposure time and hence allows for large >200x200 μm² FOVs. Further, Fig. 2C shows the obtainable FOVs for various $t_{exp}$ and a fixed line-width $w = 3\,\mu$m with our regenerative amplifier. It also shows that for 1D-LS TeFo it is more advantageous to scan the longer dimension of a given rectangular FOV than to increase the length of the shaped line. To achieve homogeneous illumination for large FOVs, it is necessary to make the line width broader ($w \sim 3 \,\mu$m) than the diffraction limit, so that subsequent pulses of the 10 kHz amplifier do overlap in the sample plane. The increased line width $w$ prevents further improvements in axial sectioning by spatial focusing when the line width approaches the diffraction limit ($w \leq 1.5 \,\mu$m; [31]). Consequently, axial sectioning is effectively independent of the scanning modality in our setup (Fig. 1D), which has the advantage of allowing for better and more direct comparison of different scanning configurations.

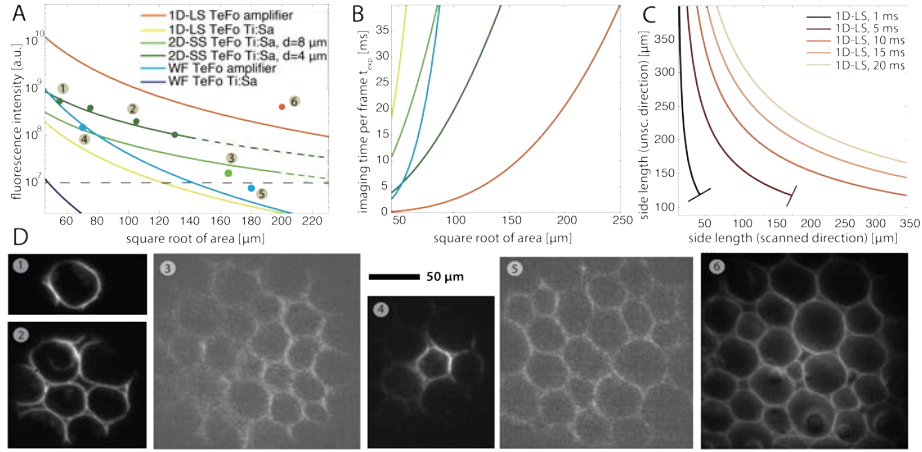

Fig. 2. Trading off area versus imaging speed and fluorescence for different excitation modalities. **A.** Theoretical estimates and experimental measurements of the fluorescence signal as a function of the excited area for different combinations of laser systems and excitation modalities. Numbers indicate parameter configurations at which experimental data were obtained with the respective images shown in **D.** Dashed lines indicate regions for which the scanning speed of the galvo mirrors would not be sufficient. The values on the y-axis are typical signal intensities for practical calcium imaging with $N_a > 1 \cdot 10^7$ representing an empirical lower bound (dashed grey line). **B.** Time required for scanning an area in order to achieve - $N_a \sim 3 \cdot 10^8$ hence sufficient signal for imaging as discussed in the main text. Coloring identical to A. **C.** Isolines for sufficient fluorescence signal ($N_a \sim 3 \cdot 10^8$) during different image exposure times in dependence on scanned and non-scanned directions for a line-scan configuration ($w \sim 3 \mu$m) using an amplifier system. **D.** Images of *convallaria* rhizome, taken with different configurations at 10% of available power and $t_{exp} = 10$ ms exposure time per plane. The scale bar applies to all subfigures.

The flexibility of our setup allowed us to realize and contrast some of these different modalities and thereby verify our theoretical findings by using a stably fluorescing, biological sample, *convallaria* rhizome (Fig. 2D). We reduced the laser power at the sample to 25 mW and 2.5 mW for the Ti:Sa and amplifier respectively (10% of the respective available maximum) to prevent overexposure and damage to the specimen. The fluorescence signal, i.e. camera count rates, match with the theoretical predictions and hence corroborate our model (circles with numbers in Fig. 2A). It is instructive to compare the amplifier WF-TeFo configuration that is very close to that used in Ref. [3] to the spiral-scan configuration using a Ti:Sa and the line-scan amplifier configuration as they result in comparable image qualities and S/B (Fig. 2D, (4), (2) and (6)). The scanning schemes therefore improve fast TeFo imaging by either lowering the requirements on the laser (2D-SS) or by increasing $A_{FOV}$ from 70 µm diameter to almost 200 µm side length. An overview of the achievable FOVs with the various scanning modalities and laser sources as well as their limitations is provided in Table 1. A summary of the possible scanning configurations and laser sources for given FOVs and frame rates is given in Table 2.

Table 1. Summary of achievable field-of-views (FOV) in fast functional imaging experiments using the various scanning modalities discussed in the main text. Note that the plane exposure time, $t_{exp}$, was fixed at 10 ms in all cases. The right column gives the main limitations when further trying to increase FOV for the various configurations. **X** denotes that in practice the laser source is not properly suited for the scanning modality.

|  | Amplifier | | Ti:Sa | | Comments |
| --- | --- | --- | --- | --- | --- |
|  | Max FOV | Limitation | Max FOV | Limitation |  |
| Wide-field (WF) | < 75 µm | Power | **X** | Power too low | No scanning required |
| Line-scan (1D-LS) | < 250 µm | Repetition rate, Optics | < 50 µm | Power | Rolling shutter readout possible |
| Spiral-scan (2D-SS) | **X** | Repetition rate | < 200 µm | Power, Galvo speed | FOV easily adaptable |

Table 2. Overview of possible scanning configuration and laser sources for given FOVs and frame rates. Nomenclature is identical to the main text. Green: expected signal intensity >$10^8$, i.e. sufficient signal for typical brain samples. Orange: expected signal levels of $10^7$-$10^8$, empirical lower bound for practical calcium imaging.

| FOV [µm] | Frame rate [Hz] | | | |
| --- | --- | --- | --- | --- |
|  | 10 | 25 | 50 | 75 |
| 50 | WF TeFo Ti:Sa<br>1D-LS Ti:Sa<br>WF TeFo Amp<br>2D-SS Ti:Sa<br>1D-LS Amp | 1D-LS Ti:Sa<br>WF TeFo Amp<br>2D-SS Ti:Sa<br>1D-LS Amp | 1D-LS Ti:Sa<br>WF TeFo Amp<br>2D-SS Ti:Sa<br>1D-LS Amp | 1D-LS Ti:Sa<br>WF TeFo Amp<br>2D-SS Ti:Sa<br>1D-LS Amp |
| 100 | 2D-SS Ti:Sa<br>1D-LS Ti:Sa<br>WF TeFo Amp<br>WF TeFo Ti:Sa<br>1D-LS Amp | 2D-SS Ti:Sa<br>1D-LS Ti:Sa<br>WF TeFo Amp<br>1D-LS Amp | 2D-SS Ti:Sa<br>1D-LS Ti:Sa<br>WF TeFo Amp<br>1D-LS Amp | 2D-SS Ti:Sa<br>1D-LS Ti:Sa<br>WF TeFo Amp<br>1D-LS Amp |
| 200 | 1D-LS Amp<br>1D-LS Ti:Sa<br>WF TeFo Amp<br>2D-SS Ti:Sa | 1D-LS Amp<br>WF TeFo Amp<br>2D-SS Ti:Sa | 1D-LS Amp<br>2D-SS Ti:Sa | 1D-LS Amp<br>2D-SS Ti:Sa |
| 300 | 1D-LS Amp<br>2D-SS Ti:Sa | 1D-LS Amp<br>2D-SS Ti:Sa | 1D-LS Amp<br>2D-SS Ti:Sa | 1D-LS Amp |

Next we verified our findings in more practical and realistic settings. In particular, we performed $Ca^{2+}$-imaging of acute brain slices of mice that expressed the $Ca^{2+}$-reporter GCaMP6m and tested all three excitation modalities (Fig. 3). Briefly, 4 month old male mice were deeply anesthetized and 1 x $10^{12}$ units of adeno associated virus (AAV) 2/8 encoding the fluorescent protein calcium sensor GCaMP6m under a neuron specific promoter (AAV2/8 phSy::GCaMP6m) were injected into the central amygdala. Three weeks after injection, transverse coronal brain slices (300 µm) were prepared. Slices were equilibrated with oxygenated ACSF for 15 min, followed by a 45 min recovery phase at room temperature, before slices were transferred to the microscope chamber.

We found that our above considerations and conclusions on fixed samples can also be applied to $Ca^{2+}$-imaging of acute brain slices. The 1D-LS TeFo configuration typically yielded a FOV >200x200 µm² at short exposure time ($t_{exp}$ = 10 ms ) also at 50 µm below the surface (Fig. 3 A). This was in contrast to the WF TeFo which even for superficial layers yielded only poor SNR for the same exposure time and smaller FOV (~75 µm) at the same sample (Fig. 3B). However, due to the experimental variability in viral infection and natural neuronal activity, the virally transfected mouse slices in general showed a broad spectrum of fluorescence intensities between different mice, slices, locations within the slice, and between cells. Fig. 3A and 3B show typical expression levels (baseline), whereas Fig. 3C shows a location in a slice of another mouse that showed particularly high fluorophore expression levels. In this region, it was possible to image with good

SNR (>10) with 2D-SS using a large (~ 200 μm diameter) FOV and an 8 μm diameter disc with $t_{exp}$ = 10 ms (Fig. 3C). Therefore we note that Fig. 3 only shows what can be typically achieved with living samples, in contrast to the demonstrations using fixed samples and theoretical estimates shown in Fig. 2. The results in Fig. 3 are therefore not meant to represent a quantitative guidance for choosing between different scanning configurations.

In our $Ca^{2+}$-imaging experiments with acute mouse brain slices, lack of appropriate stimulation equipment prevented us from invoking neuronal activity. Therefore all imaging data shows baseline fluorescence of neurons. Nevertheless, we did observed sporadic spontaneous activity in a small number of slices. An example $Ca^{2+}$-signal extracted from our imaging data is shown in Fig. 3D, where we plot the baseline fluorescence signal from an inactive neuron along with another one that showed brief and fast $Ca^{2+}$-transients. This further highlights the advantages of our high-speed TeFo techniques for fast functional imaging, demonstrating sufficient signal-to-noise and temporal resolution to record dynamic $Ca^{2+}$-signals.

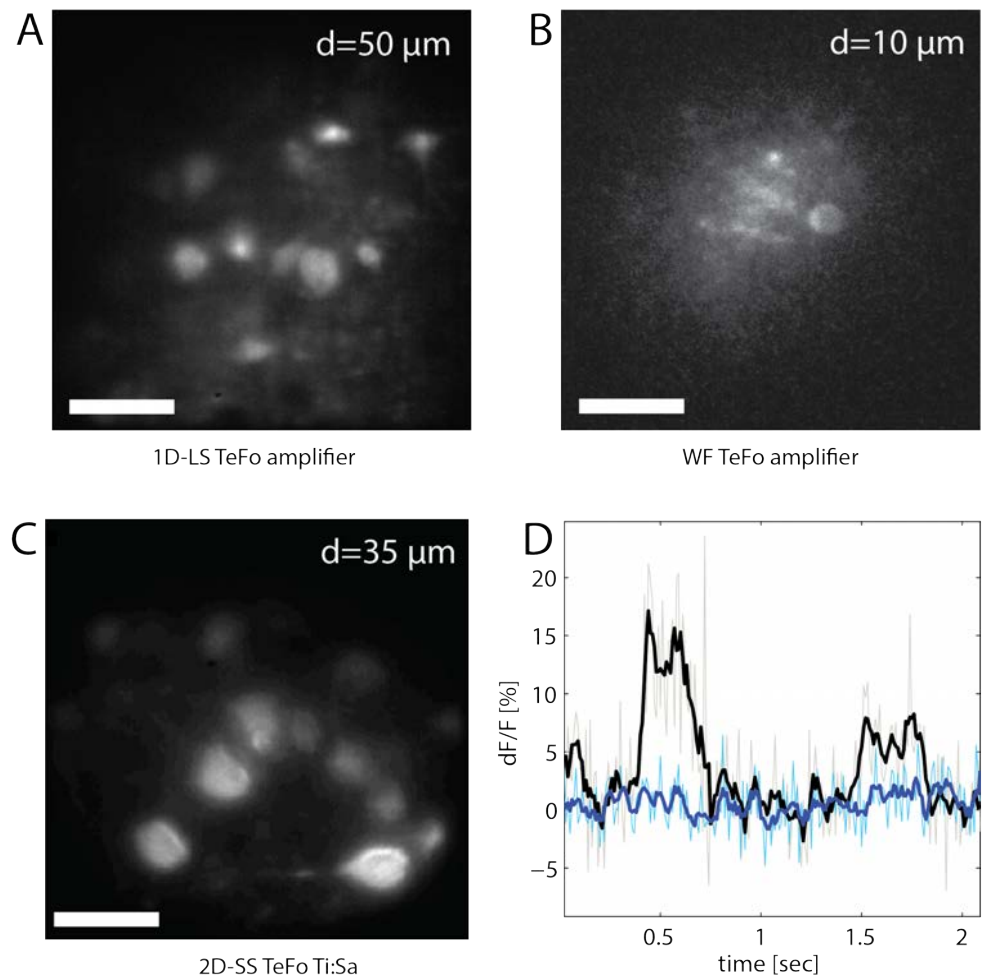

Fig. 3. $Ca^{2+}$-imaging of acute mouse brain slices at various imaging depths, and using different TeFo excitation modalities. Individual somata as well as neuropil can clearly be seen at different depths as indicated in the top right corner. **A.** Amplifier 1D-LS TeFo. **B.** Amplifier WF TeFo. **C.** Ti:Sa 2D-SS TeFo. All images were taken with an exposure time of 10 ms and no frame averaging has been performed for A-C. **D.** Extracted $Ca^{2+}$-signal from an acute brain slice that showed brief and sporadic spontaneous activity imaged with 200 μm FOV 2D-SS TeFo. The blue trace shows the baseline of an inactive neuron, while the black trace corresponds to another neuron showing brief bursts of activity. Thick lines show the 50 ms moving average of the raw extracted fluorescence signal, plotted as a relative fluorescence change dF/F. See the main text for a discussion. Scale bar is 50 μm across all images.

## 5. Suppression of scattering effects using line-scan TeFoM

Scattering of fluorescence photons in biological tissue due to refractive index inhomogeneities and cell morphologies is the limiting factor for using wide-field imaging approaches in biological tissue at larger depths. Scattered fluorescence photons lead to pixel cross-talk in the parallel (wide-field) acquisition used in all TeFo configurations discussed above. This is in contrast to point-wise two-photon scanning, where all emitted photons that are detected by a single pixel sensor contribute to the signal. It has been proposed [27] and experimentally shown [28] that scattering for scanned light sheet microscopy, a variant of line-scanning microscopy, can be reduced by utilizing the rolling shutter read-out mode of a sCMOS camera as a moving virtual 1D-pinhole, analogous to confocal imaging. In this readout mode, the detection area of the camera is restricted to a small line that moves synchronously with the illuminated line in the sample. The reduction of the active sensor area in one dimension to the width of a line allows to detect photons originating from the line-shaped illumination area on the sample only, while rejecting scattered photons (Fig. 4A). Note that in contrast to standard confocal microscopy, the purpose of this confocal slit is not to reject out-of-focus light. Since in our case illumination is done via temporal focusing excitation, which is a two-photon technique, excitation of fluorophores is already effectively reduced to the focal plane. General considerations for confocal microscopy such as axial sectioning capability [32] therefore do not apply to our discussion here. Importantly, in contrast to confocal microscopy, this scheme does not lead to any out-of-focus bleaching of fluorophores.

We applied the rolling shutter read-out approach to line-scanning temporal focusing (1D-LS TeFo). In 'free running mode', *i.e.* when the line-scan illumination and rolling shutter of the sCMOS are not synchronized, a temporally modulated intensity pattern would be observed. As an alternative to synchronization, one could use a global shutter readout mode on the sCMOS camera (if available), however this comes at the expense of a lower effective frame rate for the same exposure time. Therefore, synchronization of rolling shutter and line-scan is essential for fast imaging while at the same time it reduces scattering effects and thereby increases the achievable imaging depth.

Experimentally, synchronization of the rolling shutter of our sCMOS camera with the scanning galvo mirrors was achieved by using the camera as a master clock to trigger the galvos via a TTL signal. The scanning frequency of the synchronized system was adapted to the fixed velocity of the rolling shutter. This frequency cannot be tuned freely for currently available sCMOS cameras. In the experiments, the position of the line-shaped scanning beam in relation to the rolling shutter was aligned manually by maximizing the detected signal emitted by a fluorescent microscope slide. This adjustment is necessary to compensate for the finite response time of the galvometric scanners.

We also theoretically evaluated the expected effects of scattering reduction at different depths using Monte Carlo simulations, and validated our theoretical predictions experimentally in acute slices expressing GCaMP6m, as described above.

For the Monte Carlo simulations, we simulated fluorescence photon propagation from a point source located inside the sample to the boundary of the tissue. Then, we used ray tracing throughout the detection optics (*i.e.* objective and tube lens) onto the sensor. Simulation of photon propagation was performed similarly to Ref. [33]. In order to determine the scattering kernel, *i.e.* the scattered image of a point-like fluorophore on the sensor, we assumed photons to be emitted isotropically from a point at a given depth inside the tissue. We neglected photon absorption processes in our simulation, as these events are ~500 times less frequent than scattering events in brain tissue for the wavelength of interest [34]. The path length between two scattering events was sampled from an exponential distribution with an average scattering length $l_s = 50 \ \mu m$, corresponding to a scattering coefficient of $\mu_s = 20/mm$ for cortical tissue (grey matter) and light in the emission band of GFP and GCaMP ([34]; cf. also [35]). Scattering angles ϑ were sampled from a probability distribution based on a Henyey–Greenstein phase function [36] with scattering anisotropy $g = 0.9$ [10, 34], indicating strong forward scattering. Azimuthal angles φ were drawn from a uniform distribution between 0 and 2π. Simulations were done for $10^5$ photons, a commonly used value [33]. The scattering simulation was terminated when the surface of the tissue was reached. If the surface of the tissue was not reached after 500 scattering events, the respective photon was discarded and assumed to be absorbed. Above the sample surface, geometric ray optics was used to propagate the photons through the objective ($f = 9 \ mm$) and tube lens ($f = 200 \ mm$) of our microscope and onto the detection plane of the sCMOS camera. Photons outside the acceptance cone determined by the objective NA or the radius of the tube lens were discarded. The distance between tube lens and detector was fixed at 200 mm, and the distance between objective and tube lens was also set to 250 mm.

The pattern of light for a single point source generated by the $10^5$ simulated emissions at the detector represents the scattering kernel. The scattering kernel consists of a peak of mostly non-scattered photons and a tail that reflects the Henyey-Greenstein scattering function (Fig. 4D and inset). The fraction of non-scattered photons decreases rapidly with depth, and scattered photons result in broader distributions as imaging depth increases.

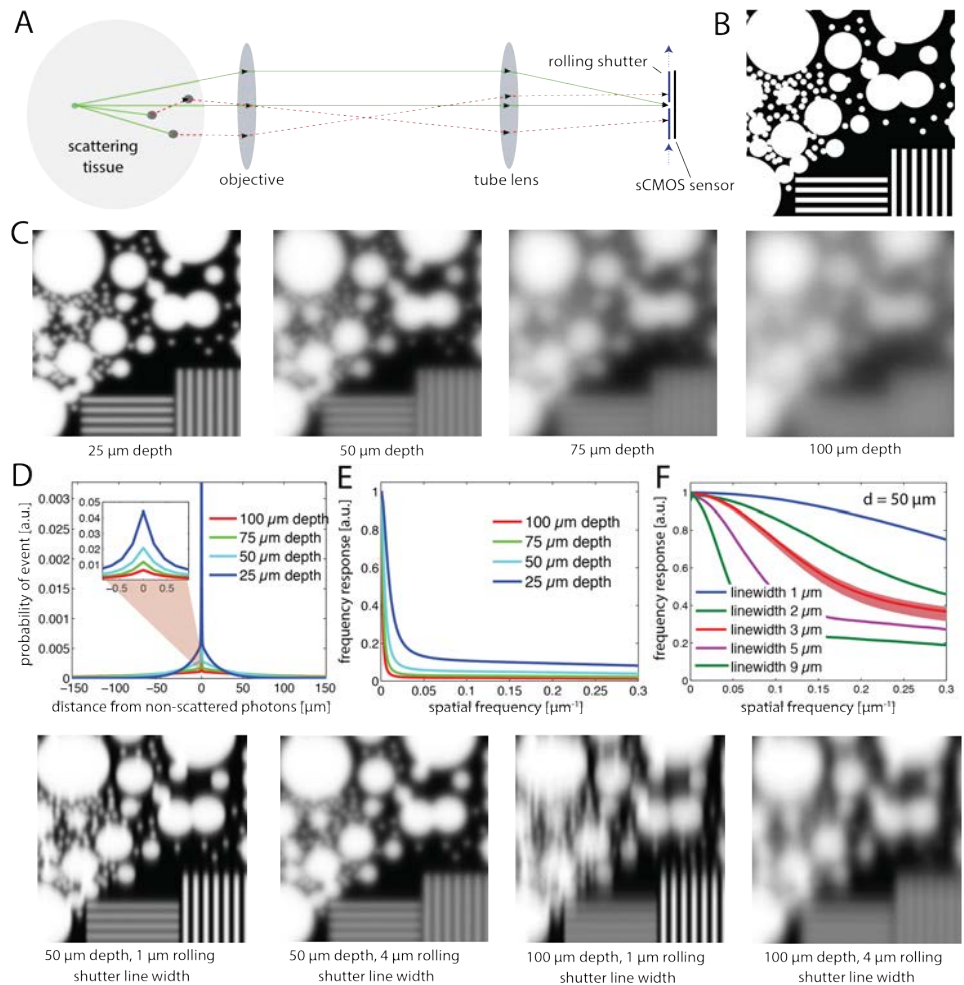

Fig. 4: Theoretical estimation of the rolling shutter effect on image quality in scattering media. **A.** Schematic drawing of the detection path. Scattered photons (dashed lines) are typically excluded by the rolling shutter. **B.** Ground truth object including circles with 30, 15, 7.5 and 2.5 μm diameter and orthogonal grids of 2 μm spacing. **C.** Simulated effect of scattering through scattering tissue ( $\mu_s = 20/\text{mm}$ ) on image quality for different imaging depths. **D.** Scattering kernel ( $\mu_s = 20/\text{mm}$ ) for different imaging depths. Note the different scaling in the inset zooms. **E.** The same scattering kernels in the spatial frequency domain. **F.** Improvement of the scattering kernel by using the rolling shutter effect for an imaging depth of 50 μm for different rolling shutter widths. The curves vary little between 25 and 100 μm imaging depth, as illustrated by the red band for the 3 μm line-width. **G.** Rolling shutter simulation of the ground truth object for different imaging depths and rolling shutter line widths.

To illustrate the effects of scattering on image quality, we used an artificial test pattern as a ground truth sample. It was composed of randomly arranged discs with 30, 15, 7.5 and 2.5 μm diameter as well as two orthogonal grids with 2 μm spacing each (Fig. 4B). The convolution of the scattering kernel with the artificial sample leads to the images in Fig. 4C. Note that this is a realistic scattering simulation only for the particular scattering length chosen here ( $l_s = 50$ μm).

For a better comparison of our theoretical and experimental results, we calculated the Fast Fourier Transform (FFT) of the scattering kernel (Fig. 4E), in which a given spatial frequency corresponds inversely to an object's size. For example, a spatial frequency of 0.1 μm$^{-1}$ corresponds to a (periodic) structure of 10 μm size. To calculate the same spatial frequency response function for detection with the rolling shutter as a moving confocal slit, a structure $W$ containing every frequency equally (*i.e.*, white noise) was randomly created as 1D image and then multiplied with the mask of a 3 μm-wide Gaussian beam $G$ to simulate the illumination. This illuminated image was then convolved with the scattering kernel $K$ (Fig. 4D) and truncated by multiplication with a mask $M$ representing the rolling shutter. To reduce the parameter space, the width of the rolling shutter was chosen to equal the illumination line width, which was found to be optimal, in accordance with Ref. [28]. This procedure was repeated for every position $x$ of the line, creating a full image $I$ by adding up the images of the single lines. Effectively, this procedure is an *in silico* reproduction of the different steps that take place when using the rolling shutter for synchronized detection: First a line in the sample is excited. Then, the emitted photons traverse the detection path and impinge on the sCMOS sensor plane on a location according to the scattering kernel. Finally, these photons are either rejected or accepted, depending on whether the mask which represents the rolling shutter is hiding the respective part of the sensor:

$$I = \sum_x M \cdot \left[ K * (G(x) \cdot W) \right] = K_{RS} * W, \qquad (2)$$

where $*$ denotes the convolution operation. From the simulation, both $I$ and $W$ are known, enabling the calculation of the effective transfer function/kernel $K_{RS}$ of the rolling shutter system using the properties of Fourier transforms (FT):

$$\hat{K}_{RS} = FT(K_{RS}) = FT(I)/FT(W) \qquad (3)$$

The resulting $\hat{K}_{RS}$ for different line- (and hence shutter-) widths and an imaging depth of 50 μm is shown in Fig. 4F and can be intuitively compared to the kernel without rolling shutter effect (Fig. 4E). The kernel changes only slightly for different imaging depths; this is shown as an example for a 3 μm wide line, where the red band in Fig. 4F shows the variability that is spanned by imaging depths from 25 to 100 μm. This reflects the fact that the resolution for a 3μm line-width is largely independent of the depth.

As it can be seen, the response for relevant structures in the range of 3-10 µm (corresponding to spatial frequencies of ca. 0.1-0.3 µm$^{-1}$) is significantly improved, especially for small line widths, compared to the scattering kernel without rolling shutter (Fig. 4E). This is also illustrated for four different parameter combinations in Fig. 4G. The spatial resolution is significantly improved, but as expected this improvement is anisotropic, and only confined to the dimension that is restricted by the rolling shutter. It is also worth mentioning that significant improvements can only be achieved for spatial frequencies smaller than the width of the illuminated line. This is an expected finding and reflects the fundamental fact that the precision of excitation determines the resolution of detection when scattering becomes dominant.

To experimentally verify the above theoretical predictions, we used our amplifier system together with the 1D-LS TeFo configuration to image GCaMP6m-labeled mouse brain slices as described above. The line width of 1D-LS TeFo was set to ~3.5 µm in order to yield homogenous excitation along the scanned direction of the plane (for $t_{exp} = 10$ ms). Rolling shutters of different widths $w_{RS}$ were used to study the obtained improvement in spatial resolution compared to the corresponding configurations without rolling shutter.

Fig. 5 shows typical results obtained in mouse brain slices at different imaging depths and with varying rolling shutter widths $w_{RS}$. A significant improvement is visible, and small structures, which were originally obscured by scatter blur, become apparent. While neuronal somata can be clearly recognized also without rolling shutter, the contrast of finer structures is clearly improved for images that use the rolling shutter synchronization. It is also obvious that small rolling shutter widths (one or two pixels, corresponding to 0.3 and 0.6 µm in the sample, respectively) cannot be used effectively because shot noise of the detection system dominates at that level due to the weak signal level. For imaging depths beyond 75 µm, out-of-focus fluorescence might additionally become relevant, since excitation becomes less confined due to scattering [10].

To compare the improvement with our theoretical results shown in Fig. 4, we computed the FFT of every image and compared the spatial frequency domains of the improved (rolling shutter) images with those where no rolling shutter was applied (Fig. 5G-I). For this comparison, it is most instructive to use the gain, $g = \log(A/B)$ with the natural logarithm, commonly measured in units of neper (Np), and the FFTs of the image with and without the rolling shutter applied, $A$ and $B$, respectively. If a certain spatial frequency is enhanced, $g$ becomes positive, and vice versa.

Two effects are clearly visible in this comparison. First, for a narrow rolling shutter width, high spatial frequencies above 0.3 µm$^{-1}$ are strongly enhanced. This corresponds to noisy pixels and partially masks our analysis in the spatial frequency domain. Second, a peak is visible for spatial frequencies below 0.2 µm$^{-1}$ (corresponding to >5 µm-sized structures). Indeed, structures of this size can be identified in rolling shutter images for 25 and 50 µm imaging depth in Fig. 5. This is in agreement with our theoretical findings, where the improvement for a 3 µm-wide line (red band in Fig. 4F) is largely restricted to spatial frequencies less or equal to ~0.2 µm$^{-1}$. The gain reaches a maximum of $g = 1$ Np which corresponds to a factor of ~2.7 for an imaging depth of 50 µm, which is however less than expected in the ideal case for a line width of 3 µm (factor of 10, $g \sim 2.3$; cf. Fig. 4E-F).

Altogether, the improvements through the rolling shutter synchronization are in line with our theoretical predictions. Thus, this approach provides an easy to implement strategy to improve image quality in scattering tissue with parallel (wide-field) acquisition modes for line-scanning excitation schemes using an sCMOS camera. To further improve upon these results for the same line width, it would be important to collect more photons and to better confine the excitation volume to reduce out-of-focus light. A simple way to collect more photons would be to lower the rolling shutter speed to achieve longer dwell time and hence collect more photons. Although currently not commercially available, a rolling shutter moving at arbitrary speeds would allow to optimize the scanning speed and dwell times depending on the scattering properties and the imaging depth.

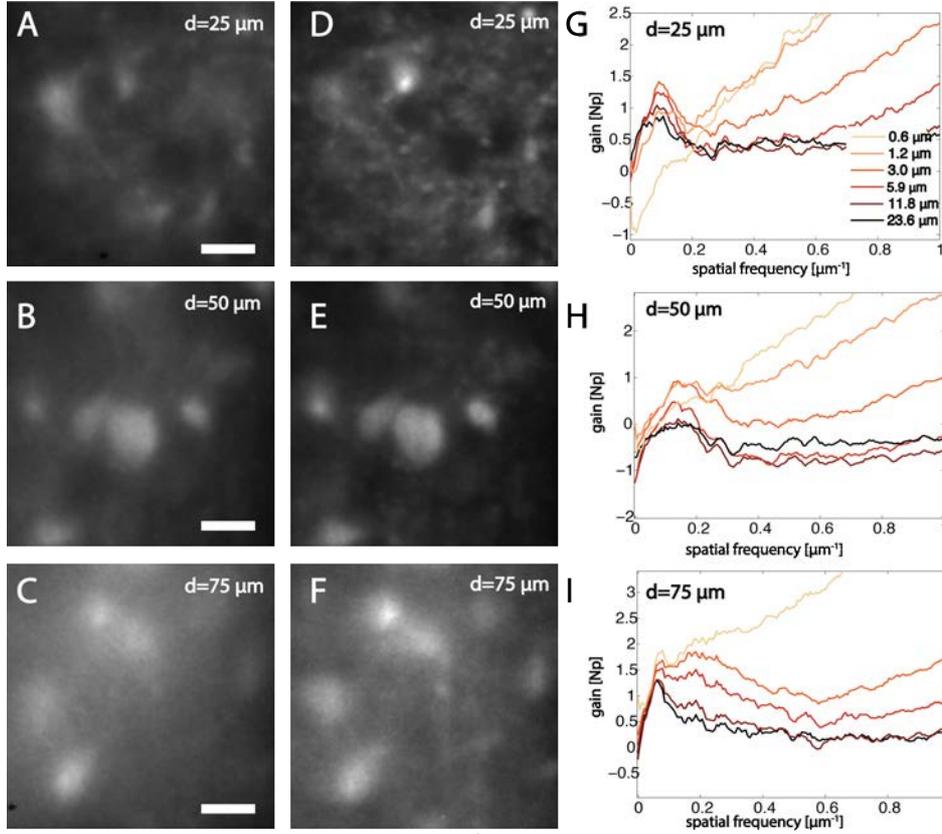

Fig. 5: Experimental demonstration of improved $Ca^{2+}$-imaging quality in scattering tissue by using line-scan TeFo illumination in combination with rolling shutter readout (sample: GCaMP6m-expressing neurons in the amygdala of acute mouse brain slices). Single images with an exposure time of 10 ms taken without rolling shutter (**A-C**) and with rolling shutter (**D-F**) at different depths (d) in the tissue. All images are an excerpt of a larger 200x200 μm FOV. **G-I.** Analysis of spatial frequency components in the acquired images. The frequency plots show which spatial frequencies are amplified in comparison with the non-rolling shutter-image, shown for different widths of the rolling shutter, $w_{RS}$. The maximum of the gain around 0.1-0.2 $\mu m^{-1}$ shows the improvement for a band of spatial frequencies corresponding to 5-10 μm sized features. Best choices for the rolling shutter width vary in dependence on imaging depth and were assessed by visual inspection: 1.2 μm for D, 3.0 μm for E and 5.9 μm for F (in I no data were acquired for $w_{RS} = 1.2 \mu m$). Scalebars are 10 μm and apply to all subfigures.

## 6. Summary and outlook

In this work we have demonstrated different strategies to improve image size, imaging speed and the image quality in functional imaging for light sculpting approaches based on temporal focusing. We have evaluated the involved trade-offs, both theoretically and experimentally. Scanning the imaging area $A_{FOV}$ in TeFo microscopy with either a line (1D-LS) or a small disc (2D-SS) instead of using the scan-less wide-field (WF) excitation such as in [3, 22] allowed us to increase the imaging field-of-view to >200x200 $\mu m^2$, while retaining a physiologically relevant imaging speed of 10 ms/plane. Spiral-scanning 2D-SS TeFo in particular can achieve an imaging FOV of ~150 μm diameter at this imaging speed using galvanometer mirrors and a standard pulsed Ti:Sa laser instead of an amplifier laser system. Additionally, spiral scanning enables simple and more flexible adjustment of the scanned area without requiring optical modifications. In this case, the FOV is limited not only by the available laser power, but also by the speed of the galvometric mirrors used for scanning. Galvo scanning speed is, however, not the limiting factor for line-scanning. Using this approach, one can increase the imaging area significantly to more than 200 μm using an amplifier laser system, while retaining high temporal resolution (10 ms per plane, translating into a frame rate of ca. 75 Hz due to limitations inherent to the sCMOS camera) using an amplifier laser system. A further increase of the image size, although possible in principle, would require specialized large aperture optics and/or lower magnification objectives. Experimentally, we have validated our theoretical findings and compared the various scanning modalities for stably fluorescing fixed samples (*convallaria* rhizome) as well as in a biologically relevant setting by performing $Ca^{2+}$-imaging at physiologically relevant time scales (ca. 75 Hz) using acute slices of mouse brain expressing genetically encoded $Ca^{2+}$-reporters.

Furthermore, we have shown that synchronizing the rolling shutter read-out mode of our sCMOS with the line-scan temporal focusing configuration can increase image quality in scattering brain tissue. We further compared our experimental findings with theoretical simulations to estimate the limits of scattering reduction using this method. The image contrast on the spatial length-scale larger or equal to the width of the illuminated line is improved considerably. Despite this, our results in deeper sections of scattering tissue such as acute mouse brain slices show that a wide-field acquisition approach using a camera falls well short of the quality obtained by point-scanning two-photon microscopy. Quantitatively, based on our results, the limit seems to reside in the range 75-100 μm, depending on the spatial extent of the structures of interest. Below this range, other approaches such as ones based on multi-photon point scanning have to be used. Therefore our parallel acquisition TeFo techniques are best suited to thin sample preparations or small model organisms with minimal scattering such as *C. elegans* or zebrafish larvae, in which the high spatio-temporal resolution outperforms standard point-scanning microscopes. To achieve further improvements in our scheme, it would be essential to collect more photons to overcome shot-noise in the detection system. This could be achieved with more efficient molecular probes or emerging sCMOS cameras featuring rolling shutters that can operate at variable speeds.

## Acknowledgments

The authors would like to thank the mechanical workshop at the IMP and the Haubensak lab members for support. This work was supported by the VIPS Program of the Austrian Federal Ministry of Science and Research and the City of Vienna as well as the European Commission (Marie Curie, FP7-PEOPLE-2011-IIF), the Vienna Science and Technology Fund (WWTF) project VRG10-11, Human Frontiers Science Program Project RGP0041/2012, Research Platform Quantum Phenomena and Nanoscale Biological Systems (QuNaBioS), and the European Community's Seventh Framework Programme (FP7/2007-2013) / ERC grant agreement n° 311701. The Institute of Molecular Pathology is funded by Boehringer Ingelheim.